\documentclass[prb,aps,twocolumn,preprintnumbers,10pt,floats,amsmath,showpacs,amssymb,superscriptaddress]{revtex4-1}
\usepackage{epsfig}
\usepackage{amsmath}
\usepackage{amssymb}
\usepackage{amsfonts}
\usepackage{mathptmx}
\usepackage{dcolumn}
\usepackage{eucal}
\usepackage{bm}
\usepackage{color}
\usepackage[colorlinks,linkcolor=blue,citecolor=blue]{hyperref}

\usepackage{epstopdf}
\usepackage{graphicx}

\begin{document}
\title{Radiation comb generation with extended Josephson junctions}
\author{P. Solinas}
\email{paolo.solinas@spin.cnr.it}
\affiliation{SPIN-CNR, Via Dodecaneso 33, 16146 Genova, Italy}
\author{R. Bosisio}
\email{riccardo.bosisio@nano.cnr.it}
\affiliation{SPIN-CNR, Via Dodecaneso 33, 16146 Genova, Italy}
\affiliation{NEST, Instituto Nanoscienze-CNR and Scuola Normale Superiore, I-56127 Pisa, Italy}
\author{F. Giazotto}
\email{giazotto@sns.it}
\affiliation{NEST, Instituto Nanoscienze-CNR and Scuola Normale Superiore, I-56127 Pisa, Italy}

\begin{abstract}
We propose the implementation of a Josephson radiation comb generator (JRCG) based on an extended Josephson junction subject to a time dependent magnetic field.
The junction critical current shows known diffraction patterns and determines the position of the critical nodes when it vanishes. When the magnetic flux passes through one of such critical nodes, the superconducting phase must undergo a $\pi$-jump to minimize the Josephson energy.
Correspondingly a voltage pulse is generated at the extremes of the junction. 
Under periodic driving this allows us to produce a comb-like voltage pulses sequence.
In the frequency domain it is possible to generate up to hundreds of harmonics of the fundamental driving frequency, thus mimicking the frequency comb used in optics and metrology.
We discuss several implementations through a rectangular, cylindrical and annular junction geometries, allowing us to generate different radiation spectra and to produce an output power up to $10$~pW at $50$~GHz for a driving frequency of $100$~MHz.

\end{abstract}
\pacs{
74.50.+r, 
74.81.Fa, 
06.20.fb, 
04.40.Nr 
}
\maketitle
\section{Introduction}
The field of optical combs has seen a growing interest in recent years \cite{udem2002optical,delhaye2007}.
The atomic clocks are extremely stable and have sharp resonances; for these reasons, they are used as time and frequency reference standards. However, their working range is limited to the radio frequency region.
This limitation has been overtaken only a decade ago. A combination of technical and conceptual improvements has allowed to extend this accuracy to higher frequency up to the optical region.
The key phenomenon is simple: the atoms are manipulated to emit periodic sharp energy
pulses which, in the frequency domain, correspond to a comb signal with the
harmonics of the fundamental frequency.
Since the generated harmonics show very sharp resonances, they can be used as a frequency standard in the optical region.
The realization of the optical frequency comb has paved the way to important applications in optical metrology \cite{hansch1999laser}, high precision spectroscopy \cite{bloembergen1977nonlinear, hansch1994frontiers} and telecommunication technologies \cite{udem2002optical, foreman2007remote}.

Recently, it has been shown that a similar frequency comb can be generated with a dc superconducting quantum interference device (SQUID) subject to a time-dependent magnetic field \cite{solinas2014josephson}.
The driving induces $\pi-$jumps of the superconducting phase which are associated to voltage pulses generated at the extremes of the device.
The voltage pulses sequence in the time domain translates, upon Fourier transforming, into a radiation comb in the frequency domain with up to hundreds of harmonics of the fundamental driving frequency.

Here, we show how similar effect can be obtained in an extended Josephson junction.
The underlying physics is similar to that discussed in Ref. \onlinecite{solinas2014josephson}, but the details of the implementation are different.
A setup involving extended junctions opens up the possibility for different geometries and, therefore, for various power spectra of the emitted radiation. In particular, since the generated radiation comb structure depends on the current-magnetic flux relation of the junction, this latter can be properly engineered in order to obtain a desired radiation power spectrum.
We discuss the rectangular, cylindrical and annular junction designs with their different strengths and weaknesses as prototypical examples of extended junctions.

The paper is structured as follows: in Section~\ref{sec:theoreticalanalysis} we introduce the physical mechanism leading to the radiation generation using extended Josephson junctions. The performance of such devices are investigated numerically in Section~\ref{sec:numericalresults}, where the emitted radiation spectra of junctions with different geometries are compared. Finally, we gather our conclusions in Section~\ref{sec:conclusions}.

%
\begin{figure}[t!]
\includegraphics[width=\columnwidth]{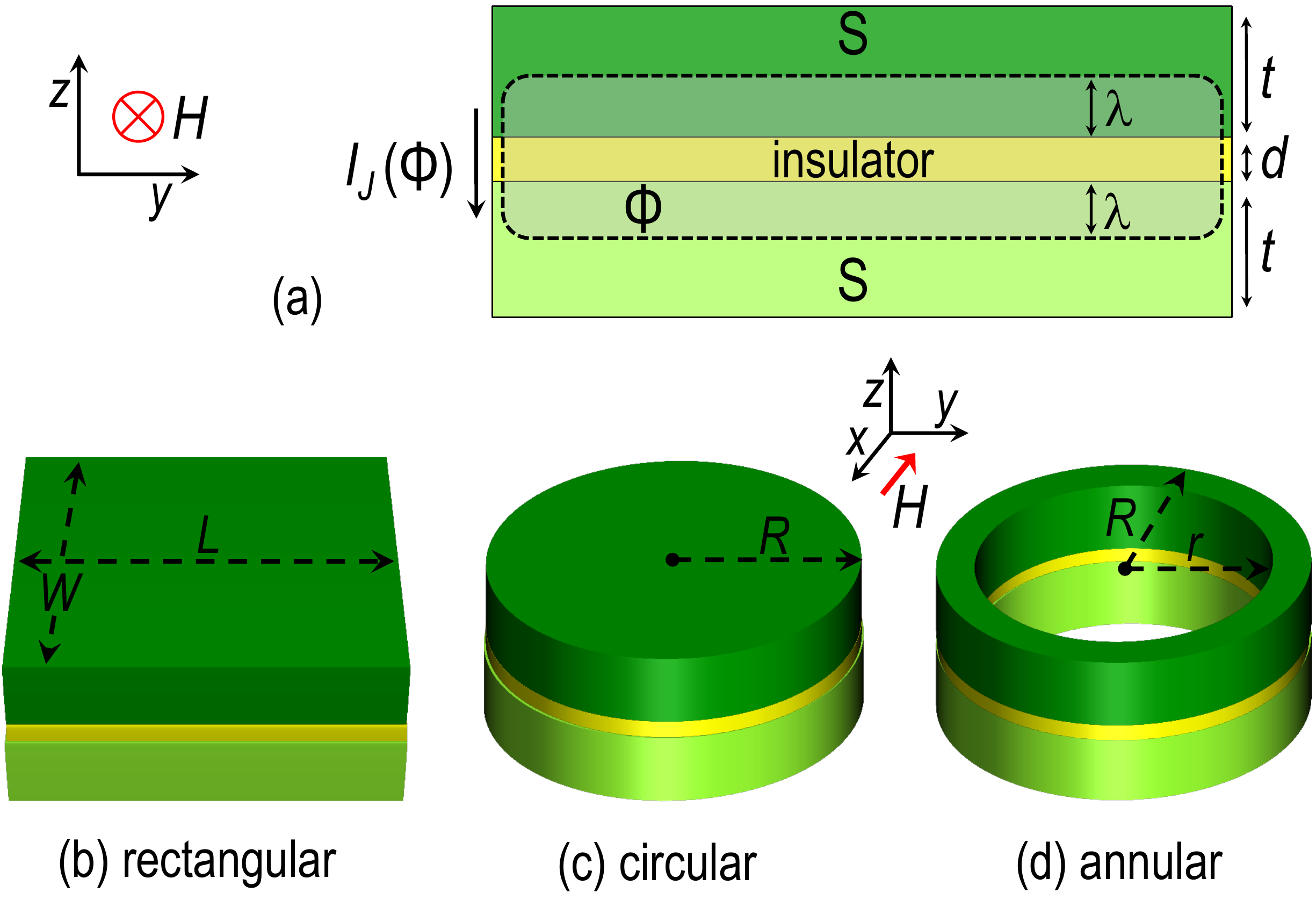}
\caption{(Color online) (a) Cross section of an extended Josephson tunnel junction in the presence of an in-plane magnetic field $H$ generating a flux $\Phi$ piercing the junction in the $x$ direction. $t$ and $\lambda$ represent the thickness and London penetration depth of the superconductors $S$, respectively, whereas $d$ is the insulator thickness. Prototypical junctions with rectangular, circular, and annular geometry are shown in (b)-(d). $L$, $W$, $R$ and $r$ are their geometrical parameters.
}
\label{fig1}
\vspace{-5mm}
\end{figure}
Our system is sketched in Fig.~\ref{fig1}(a), and consists of an extended Josephson tunnel junction composed of two identical superconducting electrodes \cite{giazotto2013coherent,martinez2013efficient,martinez2013fully,Bolmatov2010,Carty2013}.
We denote by $\lambda$ and $t$ the London penetration depth and the thickness of the superconductors $S$, respectively, which satisfy the condition $t>\lambda$. Furthermore, $d$ is the insulator thickness, whereas $t_H = 2 \lambda + d$ is the magnetic penetration thickness.

For the sake of clarity we focus on a junction in the \emph{short} limit, i.e., with lateral dimensions much smaller than the Josephson penetration depth.
In such a case the self-field generated by the Josephson current in the weak-link can be neglected with respect to the externally applied magnetic field and no traveling solitons are originated \cite{giazotto2013coherent,tinkham2012introduction}. We choose a coordinate system such that the applied magnetic field ($H$), directed along the $x$ direction, is parallel to a symmetry axis of the junction whose electrodes planes lies in the $xy$ plane [see Fig.~\ref{fig1}(a)]. 

\begin{figure}[t!]
\includegraphics[width=\columnwidth]{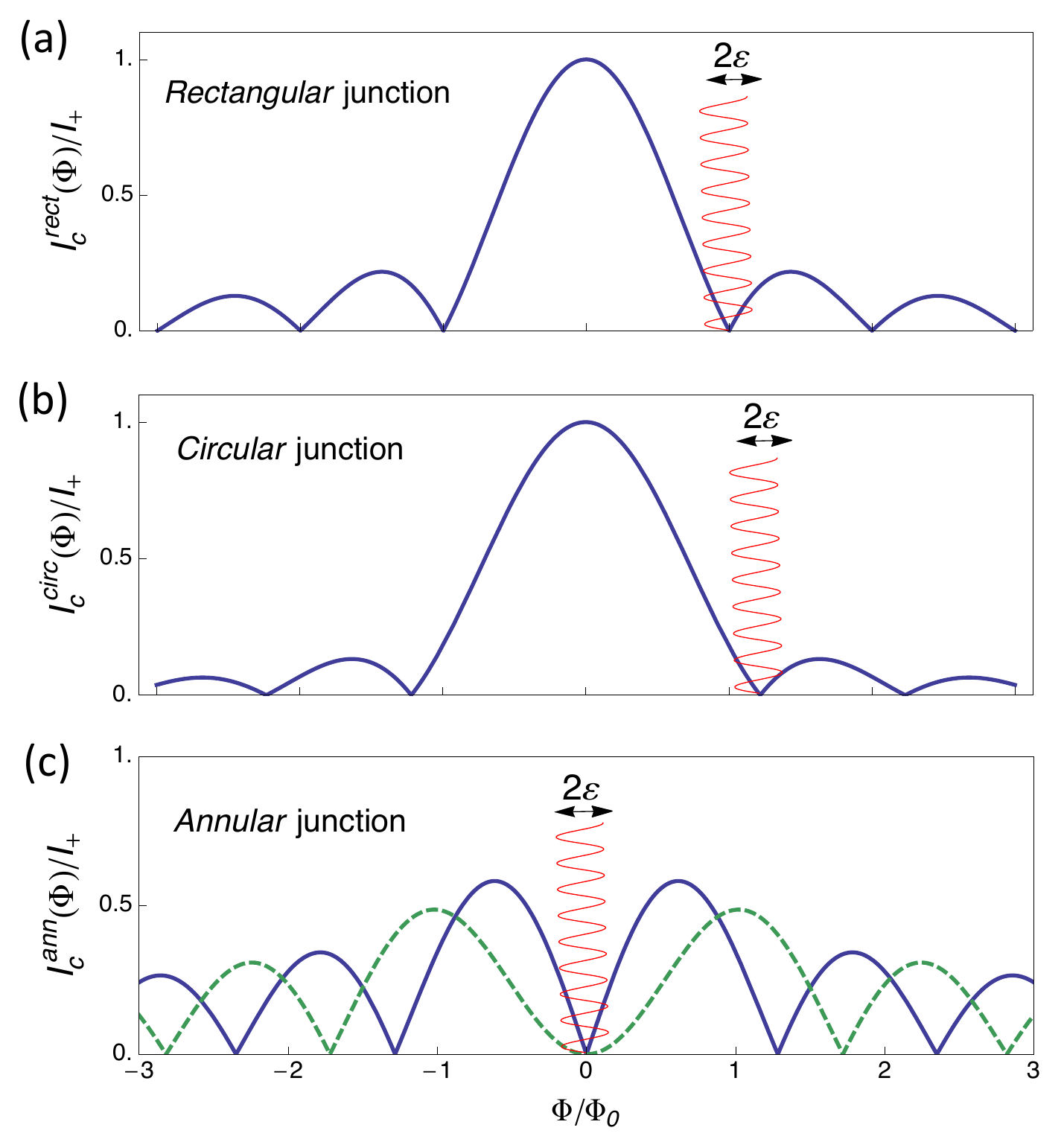}
\caption{ (Color online) Diffraction  pattern of the critical current for different geometries:
(a) rectangular junction, (b) circular junction and (c) annular junction with different numbers $n$ of trapped fluxons: $n=1$ (solid blue line) and $n=2$ (dashed green line). The red line shows $\Phi(t)/\Phi_0$ oscillating with driving frequency $\nu$ and amplitude $\varepsilon$ around different diffraction nodes.
}
\label{fig2}
\end{figure}

For the sake of presentation, we focus on a rectangular junction as in Fig.~\ref{fig1}(b), keeping in mind that a similar discussion can be extended to junctions with different geometries [Figs.~\ref{fig1}(c)-(d)].
In the limit of short junctions, the approximate behavior of the local phase is $\phi(y)= \kappa y + \varphi$ where $\varphi$ \cite{tinkham2012introduction, gross2005applied} is the superconducting phase at the center of the junction, $\kappa = 2 \pi \Phi/(\Phi_0 L)$ ($\Phi_0\simeq 2\times 10^{-15}$~Wb is  the flux quantum), $L$ is the length of the junction whereas $\Phi = \mu_0 H t_H L $ is the magnetic flux through the junction and $\mu_0$ is the vacuum permeability.
By integrating the Josephson current density per unit length \cite{tinkham2012introduction} $\mathcal{I}_c(y)$ over the junction length we obtain
\begin{equation}
 I_J^{rect}(\Phi) = \int_{-L/2}^{L/2} dy~\mathcal{I}_c(y) \sin \phi(y) =  I_+~f(t) \sin \varphi,
 \label{eq:I_J}
\end{equation}
where $f(t) = \text{sinc}(\pi \Phi/\Phi_0)$, and $I_+$ is the maximal critical current of the junction. 
The measurable critical current of the junction as a function of the magnetic flux is $I_c^{rect}(\Phi)= {\rm max}_\varphi  I_J^{rect}(\Phi)$. It displays the celebrated Fraunhofer pattern which vanishes at the diffraction nodes appearing at $\Phi= n \Phi_0$, where $n$ is an integer [Fig.~\ref{fig2}(a)].

The phase jump phenomenon we are interested in can be easily described in energetic terms.
Assuming that there is no bias current, the energy associated to the Josephson current is 
\begin{equation}
E_J (t) =\int{I_{J}^{rect} V(t)dt} = - E_{J0} f(t)\cos \varphi,
\label{eq_EJ}
\end{equation}
where $f(t)$ is the same defined above and $E_{J0}=\Phi_0 I_+/(2\pi)$. Notice that in the last step of Eq.~\eqref{eq_EJ} we have used the second Josephson relation $\dot{\varphi}=\left(2e/\hbar\right) V(t)$.
Initially $f(t=0) = 1$ and the minima of the potential energy are found at $\varphi = 2 \pi k$ (with $k$ integer).
When the magnetic flux reaches the diffraction node at $\Phi = \Phi_0$, $E_J$ vanishes, becoming negative for  $\Phi > \Phi_0$. 
To remain in a minimum energy state, $\cos \varphi$ must change sign, which implies that the superconducting phase must undergo a $\pi$ jump.
The original prediction of $\pi$ jumps \cite{giazotto2013coherent} has also been indirectly confirmed via measurements in heat transport experiments performed in temperature-biased Josephson tunnel junctions \cite{martinez2012nature,martinez2014quantum}.
Indeed, as it is explained in Ref.~\onlinecite{giazotto2013coherent}, the fact that the coherent component of the heat current remains positive when crossing a diffraction node can be understood only if the superconducting phase undergoes a $\pi$-jump.
Possible applications of this phenomenon for SQUID devices has been extensively discussed in Ref.~\onlinecite{solinas2014josephson}.

To determine the details of the voltage pulses, such as their shape and amplitude, the above discussed energetic picture is not sufficient. We rely on the so-called resistively and capacitively shunted Josephson junction (RCSJ) model \cite{tinkham2012introduction, gross2005applied} in which the Josephson junction is modelized as a circuit with a capacitor $C$, a resistor $R$, and a non-linear (Josephson) inductance $L_J$ arranged in a parallel configuration. 
We consider a sinusoidally-driven magnetic flux with frequency $\nu$ and amplitude $\epsilon$, centered in the first node of the interference pattern [Fig.~\ref{fig2}(a)], so that 
\begin{equation}
\Phi(t) = \frac{\Phi _0}{2}\,[1-\epsilon  \cos (2 \pi \nu t )]. 
\end{equation}
As a result, the magnetic flux crosses the nodes of the interference pattern at $t= (2k+1)/ 4 \nu$, with $k$ integer.
Starting from the RCSJ model we can write an equation of motion for the integrated phase $\int_{-L/2}^{L/2} dy \phi(y)$. Because of the symmetry of the problem, this reduces to a RCSJ equation for the phase at the center of the junction $\varphi$ \cite{gross2005applied}:
\begin{equation}
  \frac{\hbar C}{2 e }  \ddot{{\varphi}} + \frac{\hbar}{2 e R} \dot{\varphi}- I_+ f(t) \sin \varphi = I_B.
  \label{eq:RCSJ}
\end{equation}
We rescale the above equation in terms of adimensional time $\tau = 2 \pi \nu t$ and, using $\hbar/(2 e)= \Phi_0/ 2 \pi$, we obtain \cite{solinas2014josephson}
\begin{equation}
   c \frac{d^2 \varphi}{d \tau^2}+  \frac{d \varphi}{d \tau} - \alpha[ f(\tau) \sin \varphi - \delta]=0,
  \label{eq:RCSJ_adim}
\end{equation}
where $ \delta=I_B/I_+$, $c =2 \pi R C \nu$ and $\alpha= I_+ R/(\Phi_0 \nu)$.
The bias current is supposed to be small ($\delta \ll 1$) and its effect is to impose a preferred direction to the $\pi$ jumps of the phase. 
Furthermore, we focus on the limits $c \ll 1$ (overdamped regime) and $|I_+ R| \gg 1$, as these two conditions maximize the JRCG performance \cite{solinas2014josephson}.
%
\section{Numerical Results}
\label{sec:numericalresults}
In the following, all the numerical simulations we discuss have been performed for rectangular Nb/AlOx/Nb Josephson junctions subject to an oscillating magnetic flux with frequency $\nu=100$~MHz and amplitude $\varepsilon=0.9$.
As the junction parameters we have assumed \cite{patel1999self} $R=20~$Ohm and $I_+=0.2$~mA, while the bias current has been set to $I_B=10^{-3} I_+$. 
The relatively low frequency we have chosen, besides allowing us to achieve an important up-conversion in frequency (see Fig.~\ref{fig4} and related discussion), also assures that the small capacitance of the junction $C=10$~fF (corresponding roughly to a junction area of\cite{Likharev1986} $\simeq 0.2$ $\mu\mathrm{m}^2$) has a negligible effect on its dynamics. 

As the critical current crosses the diffraction node at $\Phi = \Phi_0$, the phase experiences a $\pi$ jump and a voltage pulse is generated across the junction.
The shape of the pulse is determined by the product $I_+ R$: the \emph{larger} $I_+ R$,  the \emph{sharper} the voltage pulse. 
Differently to what happens in the SQUID implementation \cite{solinas2014josephson}, for a rectangular junction the pulse amplitudes are not the same but show an alternating pattern of lower and higher peaks.
This is due to the asymmetry in the diffraction pattern of the critical current near the diffraction node [see Fig.~\ref{fig2}(a)].

\begin{figure}[t!]
\includegraphics[width=\columnwidth]{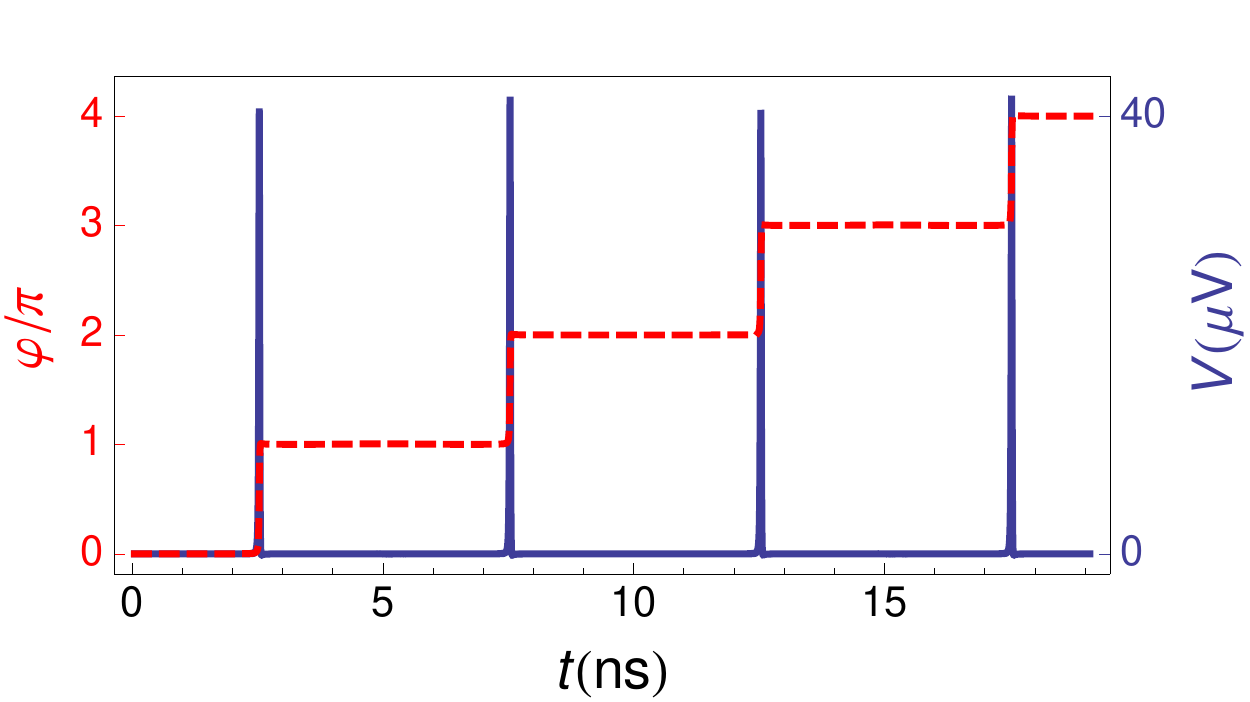}
\caption{ (Color online) Time dependence of the phase $\varphi$ (dashed line, left axis) and of the voltage $V$ (solid line, right axis) across a rectangular Josephson junction.
One-directional jumps can be realized by applying a suitable $I_B$. The voltage pulses are generated when the interference node is crossed. The parameters chosen for the calculations are those typical of a Nb/AlOx/Nb junction \cite{patel1999self} with $R=20$~Ohm, $I_+ = 0.2$ mA, and we set $\epsilon=0.9$. The typical junction capacitance is about $C= 10$ fF and it has been neglected as it does not affect the junction dynamics.
}
\label{fig3}
\vspace{-6mm}
\end{figure}

\begin{figure}[t!]
\includegraphics[width=\columnwidth]{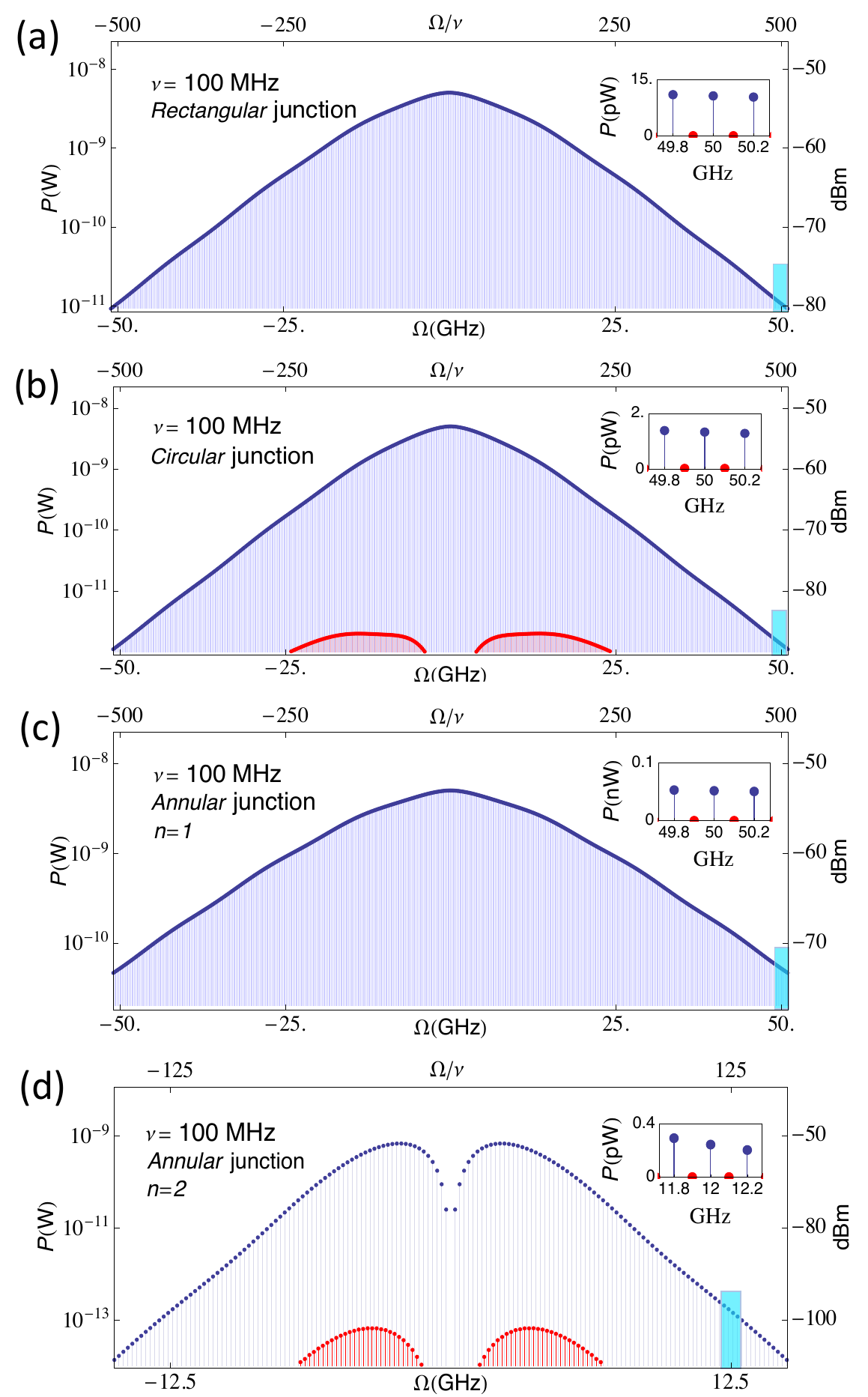}
\caption{(Color online) Power spectrum of the Josephson radiation comb generator over a $50$~Ohm transmission line.
The output power is shown in W (left) and dBm (decibel-milliwatt) (right).
The calculation is performed for a $N=10^3$ chain of Nb/AlOx/Nb extended Josephson junctions subject to a $\nu = 100~$MHz driving. The cyan regions correspond to the insets.
The panels refer to different geometries: (a) rectangular junction, (b) cylindrical junction, (c) annular junction with one fluxon trapped and (d) annular junction with two fluxons trapped. 
Blue and red points indicate the even and odd harmonics, respectively.
In Figs. (a) and (c) only the \emph{even} harmonics are present because of the comb-like shape of the voltage pulses.
The junction parameters were set as those in Fig. \ref{fig3}.
}
\label{fig4}
\end{figure}

This real-time voltage comb could be used in several ways.
One is as a generator of equally spaced voltage pulses to be used in electronics \cite{levitons}.
A second one is as a high precision controller.
In fact, because of the Josephson relation, the time average of a voltage pulse is actually quantized as a consequence of the $\pi$ jump of the phase: 
\begin{equation}
\frac{2 e}{\hbar}\, \int_{t_1}^{t_2} dt V(t) = \Delta \varphi = \pi,
\end{equation}
where $t_1$ and $t_2$ are the times in which the jump begins and ends, respectively.
The phase jumps do not depend on the dynamics or on the speed of the node crossing.
The only condition to be satisfied is the crossing of the diffraction node.
This makes the pulse generation robust against imperfection in the dynamics of the junction and the driving.
A possible application could be the high precise control of a quantum logic gate for superconducting based qubits \cite{nakamura1999,makhlin2001}.
\\
The voltage pulse sequence in Fig.~\ref{fig3} has even more interesting applications as a radiation generator.
In fact, in the frequency domain it corresponds to a frequency comb similar to the ones used in optics \cite{udem2002optical}. 
To test this possible implementation we have calculated the power spectrum $P$ vs frequency $\Omega$.
We first compute the Fourier transform of the voltage 
\begin{equation}
V(\Omega) = \int_{0}^T dt e^{i \Omega t} V(t).
\end{equation}
The power spectral density (PSD) is then 
\begin{equation}
\text{PSD}(\Omega) = \frac{1}{T}\, | V(\Omega)|^2.
\end{equation}
Finally, the power $P$ is calculated by integrating the PSD around the resonances $k \nu$ (where $\nu$ is the monochromatic driving frequency) and dividing for a standard load resistance $R_L$ of $50$~Ohm. 
This is the power we would measure at a given resonance frequency with a bandwidth exceeding the linewidth of the resonance.\\
\indent To increase the output power, we have considered a linear array of $N$ identical junctions connected together via a superconducting wire as done for the metrological standard for voltage based on the Josephson effect \cite{shapiro1963josephson, kautz1987precision,tsai1983high, solinas2014josephson}.
The coupling among different junctions has been neglected: This condition can be realized in practice by a suitable design choice which reduces the cross capacitance and the inductance between neighbor junctions.
In this case the current conservation through any $i$-th junction leads immediately to a set of
decoupled RCSJ equations of the form~\eqref{eq:RCSJ_adim} \cite{solinas2014josephson}.
Therefore, the dynamics of the junctions are independent and the voltage at the extremes of the
array is found by summing up the voltages of the single junctions.
Under the hypothesis that all the junctions in the chain generate the same voltage $V(t)$, the total voltage drop across the device is simply $V_\text{tot}(t)=N\,V(t)$. Accordingly the \emph{intrinsic} power, that is, the power delivered to an ideal load, would scale as $N^2$. On the other hand, the \emph{extrinsic} power is less trivial and depends on the detection system used\footnote{For instance, with a simple detection system as the one discussed in Ref.~\onlinecite{solinas2014josephson}, it was shown that under certain conditions the power may scale linearly with the number $N$ of elements of the array, rather than $N^2$.}. 
The $N^2$-scaling of Josephson junctions performance is also a widely studied topic in the context of phase-locked Josephson junctions arrays \cite{Jain1984,Barbara1999,Ozyuzer2007}.
In order to get an estimate of the device performance, in the following we have calculated the emitted power for arrays of $N=10^3$ junctions by dividing the total voltage by a standard load resistance $R_L=50$~Ohm as mentioned above.

Limitations to this simplified analysis can arise if we must take into account the effects of propagation of the emitted radiation along the chain \cite{solinas2014josephson}. In fact, in our model we have considered the device as a lumped element and this assumption breaks down as the length of the chain approaches the wavelength of the emitted radiation.
As a quantitative estimate, the minimum wavelength $\lambda_{min}$ (emitted at 50 GHz) must satisfy the relation $2 \lambda_{min} \geq L$ where $L$ is the total length of the device \cite{solinas2014josephson}.
Considering a packing density of the junctions of $5~\mu$m, the above condition is satisfied for $N=10^3$ junctions.
Despite being detrimental for the device performance, the propagation effects can be taken into account and corrected. 
With a careful design of the device, they could also be exploited to amplify the output power at specific working frequencies.
Another limiting factor may be an intrinsic property of the device, such as the flux flow through the superconductor \cite{Vanacken2000}. This could perturb the external magnetic flux that we use to induce the $\pi$-jumps of the phase. Closely related parasitic effects have also been studied recently \cite{Bosisio2015parasitic} for a similar system based on SQUIDs. Despite being beyond the scope of the present work, this remains an interesting issue that would require further investigation.

Figure~\ref{fig4}(a) shows the emitted radiation power spectrum for a chain of $N=10^3$ Nb/AlOx/Nb rectangular junctions\cite{gross2005applied,lloyd1987,popel19901, patel1999self} driven by a $100~$MHz oscillating magnetic field.
As we can see, the device is able to provide a power of about 10 p$W$ at $50~$GHz (corresponding to the $500$-th harmonic of the driving frequency). 

The implementation with extended junctions opens the way also for a geometric optimization. 
Choosing a different junction geometry affects the critical current of the junction and, therefore, the position and the form of the nodes. 
For the circular geometry the Josephson current exhibits the known Airy diffraction pattern, 
\begin{equation}
I_J^{circ}(\Phi)=I_+ \frac{J_1(\pi\Phi/\Phi_0)}{(\pi\Phi/2\Phi_0)}\sin \varphi,
\end{equation} 
where $J_1(y)$ is the Bessel function of the first kind, $\Phi=2\mu_0 HRt_H$ and $R$ is the junction radius.
For the annular junction \cite{Martucciello1996,Nappi1997}, the Josephson current takes the form
\begin{equation}
I_J^{ann}(\Phi)=I_+\frac{2}{1-\alpha^2} \int^1_\alpha dx\,xJ_n(x\pi \Phi/\Phi_0)\sin \varphi, 
\end{equation}
where $\Phi=2\mu_0 HRt_H$, $\alpha=r/R$, $J_n(y)$ is the $n$th Bessel function of integer order, $R$ ($r$) is the external (internal)  radius, and $n=0,1,2,...$ is the number of trapped fluxons in the junction barrier.
The critical currents for these two junction geometries are defined as $I_c^{circ}(\Phi)= {\rm max}_\varphi  I_J^{circ}(\Phi)$ and $I_c^{ann}(\Phi)= {\rm max}_\varphi  I_J^{ann}(\Phi)$, respectively, and they are shown in Fig.~\ref{fig2} (b) and (c), respectively.
We note that the position of the diffraction nodes is different from the rectangular geometry case: The critical currents vanish for non-integer values of the ratio $\Phi/\Phi_0$. Moreover, in the annular case, we see that the slope of $I_c^{ann}(\Phi)$ at $\Phi=0$ differs if the number of fluxons $n$ changes. The parameter $n$ can thus be seen as an additional degree of freedom which has important effects on the shape of the emitted radiation power spectrum [see Figs.~\ref{fig4}(c) and~(d)].
The driving is assumed to have the same periodic behavior, oscillating around the diffraction nodes as shown in Fig.~\ref{fig2}.

The power spectrum of the circular junction [Fig.~\ref{fig4}(b)] is similar to the rectangular junction one.
It generates smaller output power at high frequency reaching $2~$pW at $50~$GHz.

Particularly relevant is the annular junction case. Here, we have an additional controllable parameter: the number $n$ of fluxons trapped in the junction.
The most interesting situation is when there is one fluxon trapped ($n=1$) [see Fig.~\ref{fig4}(c)].
In this case, it is possible to modulate the magnetic flux near the vanishing point [see Fig.~\ref{fig2}(c)] making the flux driving easier.
In addition, the diffraction pattern is highly symmetric near $\Phi=0$.
This allows one to generate a very precise voltage pulse patterns that is eventually reflected in a stronger power output at high frequency, as shown in Fig.~\ref{fig4}(c) ($0.1$~nW at $50$~GHz).

By varying the number of fluxons, the junction diffraction pattern changes [see Fig.~\ref{fig2}(c)]: Correspondingly, the dynamics of the junction is different, generating different emitted radiation power spectra.
Figure~\ref{fig4}(d) shows the spectrum generated by an annular junction chain when two fluxons are trapped in each junction.
The overall power emitted is smaller and the signal is accessible up to $\sim 10$~GHz.
The spectral features are very different with respect to the single fluxon ones [Fig.~\ref{fig4}(c)].
In particular, the lower harmonics (a few multiple of $\nu$) are now suppressed while the output maximum arises around a few GHz.

The insets in Fig.~\ref{fig4} represent the blow-up of the cyan regions in the main panels: Notice that since the voltage pulse signals are almost rectified (Fig.~\ref{fig3}), the spectra contain predominantly the even harmonics, the odd ones being orders of magnitude smaller.

The main sources of error that can limit the device performance are the imprecisions in the fabrication process.
Small differences in the geometry of the junctions, i.e., length $L$ for the rectangular junctions and radii $R$ and $r$ for the circular and annular junctions, will produce off-sets in the fluxes and delays in the phase jumps.
The voltages will still sum up but the total voltage pulse shape will be broadened by these effects.
In the frequency domain, this corresponds to an additional cut-off at high frequency. 
Another potential detrimental factor is the correction to the dynamics due to the intrinsic junction capacitance.
However, this effect can be accounted for, minimized or corrected by a proper device design.

Furthermore, notice that our whole description is done by considering the effect of the time-dependent magnetic field only, assuming there is no induced electric field on the junctions. It is well known that oscillating magnetic field can generate in turn electric fields which, especially at high frequency ($\sim$GHz), may significantly alter the effect we discuss. However, this problem can be overcome by embedding the junction chain inside a suitably designed cavity where the electric (TE) and magnetic (TM) modes are spatially separated\cite{Eaton2010,Goryachev2015}.

The junction array configurations discussed above are suitable for the use as radiation emitters up to $50$~GHz. 
The most straightforward way to detect the power in this frequency range is to couple the device to a transmission line and to feed the signal to a commercial spectrum analyzer.
To have access to higher frequency we must use different materials (for example, YBCO as discussed in Ref.~\onlinecite{solinas2014josephson}) or adopt specific chain design.
This change must be accompanied with a new detection schemes, for example, by using antennas coupled to the device electrodes \cite{solinas2014josephson}.
Finally, in light of possible implementations, besides the Nb/AlOx/Nb junctions considered in the present work, we signal that other materials could be promising candidates. For instance Nb/HfTi/Nb junctions \cite{Niemeyer2002,Koelle2013}, being SNS-like (superconductor-normal metal-superconductor), would have the advantage of having almost negligible capacitance, despite having a slightly lower $I_+ R$ product.
\section{Conclusions}
\label{sec:conclusions}
In summary, we have discussed the possibility to realize a Josephson radiation comb generator with \emph{extended} Josephson junctions driven by a time-dependent magnetic field.
With a linear array of $N=10^3$ Nb/AlOx/Nb junctions and a driving frequency of $100~$MHz, we estimate that substantial power [up to $\sim$100~pW] can be generated at $50~$GHz ($500$-th harmonics), opening the way to a number of applications.
The device has room for optimization by modeling the geometry of the single junctions, the fabrication materials (see, for example, Ref.~\onlinecite{solinas2014josephson}), the driving signal and the array design.
The discussed implementation would have the advantage to be built on-chip and integrated in low-temperature superconducting microwave electronics.
\section{Acknowledgments}
Stimulating discussions with C. Altimiras, S. Gasparinetti and D. Golubev are gratefully acknowledged.
P.S. has received funding from the European Union FP7/2007-2013 under REA
Grant agreement No. 630925 -- COHEAT and from MIUR-FIRB2013 -- Project Coca (Grant
No.~RBFR1379UX). The work of  R.B. has been supported by MIUR-FIRB2013 -- Project Coca (Grant
No.~RBFR1379UX).
F.G. acknowledges the European Research Council under the European Union's Seventh Framework Program (FP7/2007-2013)/ERC Grant agreement No. 615187-COMANCHE for partial financial support.


\end{document}